\documentclass[]{proarticle}
\usepackage{multicol}
\usepackage{graphicx}
\usepackage{amssymb}
\usepackage{epsfig}

\def\bea{\begin{eqnarray}}
\def\eea{\end{eqnarray}}

\usepackage{amsthm,bm,mathrsfs,stmaryrd}
\usepackage{slashed,tikz}
\usepackage{enumerate}
\usepackage{youngtab,simplewick}

\usepackage{cite} 
\usepackage{longtable}

\newcommand{\tfrac}[2]{{\textstyle\frac{#1}{#2}}}
\newcommand{\eqref}[1]{(\ref{#1})}

\newcommand{\be}{\begin{equation}}
\newcommand{\ee}{\end{equation}}
\newcommand{\ba}{\begin{eqnarray}}
\newcommand{\ea}{\end{eqnarray}}

\newcommand{\mt}[1]{$\mathop{#1}$}

\newcommand{\sst}{\scriptscriptstyle}

\newcommand{\nn}{\nonumber\\}

\begin{document}
\title{On cubic AdS interactions of mixed-symmetry
 higher spins}
\author{L. Lopez}
\maketitle
\address{Scuola Normale Superiore and INFN\\
Piazza dei Cavalieri 7, 56126 Pisa, Italy}
\eads{luca.lopez@sns.it} 
\begin{abstract}
The problem of finding consistent cubic AdS interactions of massless mixed-symmetry higher-spin fields is recast into a system of  partial differential equations that can be solved for given spins of the particles entering the cubic vertices. For simplicity, we consider fields with two families of indices for which some examples of interactions are explicitly discussed.
\end{abstract}


\begin{multicols}{2}
\section{Introduction}

Understanding the relation between String Theory (ST) and higher-spin (HS) field theories is currently regarded as a key step in order to gain new insights into both.\footnote{See \cite{Bekaert:2010hw,Sagnotti:2011qp} for recent reviews on HS fields.} 
One of the main obstacles in pursuing this goal is our limited knowledge of the behavior of mixed-symmetry fields\footnote{See \cite{Campoleoni:2008jq} and references therein for mixed-symmetry fields in flat space.} in AdS background. These types of fields naturally appear in ST and are believed to underlie most of its extraordinary features. Although a consistent theory of interacting  massless symmetric HS fields has been known for a long time \cite{Vasiliev:1988sa}, when it comes to mixed-symmetry fields the situation is completely different. So much so that an understanding of the free theory has been attained only very recently \cite{Zinoviev:2002ye,deMedeiros:2003px,Burdik:2011cb,Alkalaev:2005kw,Zinoviev:2009vy,Alkalaev:2010af,Skvortsov:2009zu,Alkalaev:2009vm,Grigoriev:2012xg,Campoleoni:2012th}, while the interactions problem have been little investigated so far \cite{Alkalaev:2010af,Zinoviev:2010av,Boulanger:2011qt,Zinoviev:2011fv,Boulanger:2011se}.

The key differences between flat and AdS backgrounds show up as soon as massless mixed-symmetry fields are considered. Indeed, while unitary gauge fields in flat space  possess one gauge parameter for each rectangular block present in the corresponding Young tableaux (YT), their AdS counterparts only possess the gauge parameter whose diagram lacks one box in the upper rectangular block \cite{Metsaev:1995re}. As result, in general an AdS massless field propagates more degrees of freedom than the corresponding flat-space one. The exact pattern of flat-space gauge fields associated to a single AdS one was first conjectured by Brink, Metsaev and Vasiliev (BMV) in \cite{Brink:2000ag} and was then proved in \cite{Boulanger:2008up,Alkalaev:2009vm}

These notes are aimed at setting the stage for a systematic study of the cubic interactions of massless mixed-symmetry fields in AdS. For simplicity, we restrict our attention to fields with two families of indices which describe the most general representations of the AdS isometry group \mt{SO(d-1,2)} up to \mt{d=6}\,. Moreover, we focus on those portions of the couplings that do not contain divergences, traces or auxiliary fields (here denoted by TT parts for brevity). The reason is that, besides being simpler to determine, they suffice to encode the on-shell content of the theory. Relying on the Noether procedure in the ambient-space formalism, we show that this problem is equivalent to finding polynomial solutions of a set of linear partial differential equations (PDEs), along the lines of \cite{Taronna:2011kt,Joung:2011ww,Joung:2012rv}. Although rather complicated, the latter can be explicitly solved for given spins of the particles entering the cubic vertices and we describe in some detail the interactions between the simplest hook gauge field and two spin-1 gauge fields as an example. The resulting AdS interactions present a highest-derivative part dressed by a chain of lower-derivative vertices size by corresponding powers of the cosmological constant. This structure is very similar to the one found by Fradkin and Vasiliev (FV) \cite{Fradkin:1986qy} for the gravitational interactions of  totally-symmetric HS fields in AdS. In the FV system, the highest-derivative part has exactly the same form as the flat-space vertices which can be singled out considering a proper (non-singular) flat limit \cite{Boulanger:2008tg}. However, as far as mixed-symmetry fields are concerned, the situation is in principle subtler due to the non-conservation of degrees of freedom. A correct analysis requires the introduction of appropriate St\"{u}ckelberg fields according to the BMV pattern in order to guarantee that the same amount of gauge symmetry be present on both sides. We postpone this issue for a future work.  
 
The organization of this paper is the following: Section \ref{sec2} is devoted to the formulation of the free theory of massless mixed-symmetry HS fields in the ambient-space formalism. In Section \ref{sec3} we show how to recast the consistency condition for the TT parts of the interactions into a system of PDEs. The entire discussion focuses on the two-family case. Finally, we solve the system describing the interactions between the simplest hook gauge field and two spin-1 gauge fields.

%
\section{Ambient-space formalism for HS fields}\label{sec2}
%
The ambient-space formalism \cite{Fronsdal:1978vb,Biswas:2002nk} regards 
the \mt{d}-dimensional AdS space as the codimension-one hypersurface
 \mt{X^{2}=-L^{2}} embedded into a \mt{(d+1)}-dimensional flat-space.
In our convention the ambient metric
is \mt{\eta=(-,+,\,\ldots,+)}, so that 
AdS is actually Euclidean. 

Focusing on the region \mt{X^{2}<0}\,, there exists an isomorphism between multi-symmetric tensor fields in AdS, \mt{\varphi_{\mu_{1}\cdots\mu_{s_{1}};
\cdots;\,	\rho_{1}\cdots\rho_{s_{\kappa}}}}\,, and corresponding ambient-space fields, \mt{\Phi_{M_{1}\cdots M_{s_{1}};
\cdots
	;\,P_{1}\cdots P_{s_{\kappa}}}}\,, that satisfy
	the \emph{homogeneity} and \emph{tangentiality} (HT) conditions
\ba
	&\mbox{\rm Homogeneity}: 
	&(X\cdot\partial_{X}-\Delta)\,\Phi(X,U)=0\,,
	\label{h con} \\
	&\mbox{\rm Tangentiality}: 
	&X\cdot \partial_{U^{m}}\,\Phi(X,U)=0\,.	
	\label{t con}
\ea
Here, \mt{m=1,\ldots,\kappa} runs over the number of independent families of symmetrized indices, while the degree of homogeneity \mt{\Delta} parametrizes the value of the quadratic Casimir, \emph{i.e.} the AdS mass. As it will prove very convenient in describing the interactions, we have introduced the auxiliary-variable notation for the 
ambient-space fields
\ba
\label{gen mix}
	&& \Phi(X,U)= \frac{1}{s_{1}!\cdots s_{\kappa}!}\,\Phi_{M_{1}\cdots M_{s_{1}};
	\cdots;\,
	P_{1}\cdots P_{s_{\kappa}}}(X)\nn
	&&\times\, 
	({U}^{\sst 1})^{M_{1}}\cdots({U}^{\sst 1})^{M_{s_{1}}}
\cdots 
({U}^{\sst \kappa})^{P_{1}}\cdots({U}^{\sst \kappa})^{P_{s_{\kappa}}},
	\ea
where \mt{s_{1}\ge \cdots \ge s_{\kappa}}\,.	
Since no definite symmetry properties between different families of indices are enforced, these fields define reducible tensors.

In the following we restrict the attention to massless fields, namely to short representations associated with gauge symmetries\footnote{Requiring constraints on the gauge parameter that are less stringent than \eqref{t con E} other symmetries would be allowed \cite{Alkalaev:2009vm,Joung:2012rv}. These are associated to (non-unitary) partially-massless fields \cite{Deser:1983mm,Deser:2001pe,Skvortsov:2006at,Francia:2008hd} that we will not consider in this paper.} \cite{Metsaev:1995re} 
\ba
\label{gt}
&&\delta^{\sst (0)}\,\Phi(X,U)=U^{{m}}\!\cdot\partial_X\,E^{{m}}(X,U)\,,\\
\label{t con E}
&&X\cdot \partial_{U^{m}}\,E^n(X,U)=0\,.
\ea
However, as we have anticipated not all these reducible gauge symmetries can be preserved in AdS. Indeed, the compatibility with the HT conditions (\ref{h con}\,,\,\ref{t con}) 
restricts both the possible values
of \mt{\Delta}\, and the irreducible components of \mt{E^m}. More precisely, it translates into the following set of equations
\ba
\label{h con E} 
	(X\cdot\partial_{X}-\Delta-1)\,E^{{m}}(X,U)&=&0\,,\\
	(\Delta+1)\,E^{{m}}(X,U)-	U^{{l}}\!\cdot\partial_{U^{m}}\,E^{l}(X,U)&=&0\,.\label{constr E}
\ea	
Henceforth we focus on the the two-family case, for which the system \eqref{constr E} becomes
\ba
\label{2famC}
&(\Delta+2-s_1)\,E^1=U^2\!\cdot\partial_{U^1}\,E^2\,,\\
\label{2famC2}
&(\Delta+2-s_2)\,E^2=U^1\!\cdot\partial_{U^2}\,E^1\,.
\ea
Let us analyze separately two cases:\\
{\bf 1)} \mt{\Delta\neq s_1-2}\,. In this case, one can substitute eq.~\eqref{2famC} into  eq.~\eqref{2famC2}, ending up with 
\be
\label{E1cond}
(\Delta+2-s_1)\,(\Delta+2-s_2)\,E^2=U^1\!\cdot\partial_{U^2}\,U^2\!\cdot\partial_{U^1}\,E^2\,.
\ee
At this point, one can decompose the reducible gauge parameter \mt{E^2} in terms of its irreducible components as
\ba
\label{irrexp}
&&E^2=\sum_{n=0}^{s_2\!-\!1}\left(U^2\!\cdot\partial_{U^1}\right)^{s_2\!-\!1-\!n}\,\hat{E}^2_{\{s_1\!+\!s_2\!-\!1\!-\!n,n\}}\,,\\
&&U^{{1}}\!\cdot\partial_{U^{2}}\,\hat{E}^2_{\{s_1\!+\!s_2\!-\!1\!-\!n,n\}}=0\,,
\ea
where the labels \mt{\{s_1,s_2\}} identify the structure of the corresponding YT 
\be
\label{yng}
\overbrace{\yng(8,6)}^{s_{1}}
\put(-92,0){$\underbrace{\hspace*{68pt}}_{s_{2}}$}.
\ee
Plugging the expansion \eqref{irrexp} into eq.~\eqref{E1cond}, all the irreducible components, except the ones with \mt{n} satisfying    
\be
\label{nequ}
(\Delta+2-s_1)\,(\Delta+2-s_2)=(s_1-n)\,(s_2-n)\,,
\ee
identically vanish. If \mt{\Delta} is not an integer, this equation does not admit any solution for \mt{n}\,. As a consequence, no gauge symmetry is allowed and the corresponding fields are massive.\footnote{Here by massive fields we mean the ones that do not possess any gauge symmetry. Partially-massless symmetries that we  have mentioned above can only appear for integer values of \mt{\Delta}\,.} On the other hand, assuming \mt{\Delta\in\mathbb Z}\,, the solutions to eq.~\eqref{nequ} together with the corresponding gauge parameters are given by 
\begin{itemize}
\item \mt{n=\Delta+2}  
\be
\label{n1}
E^2=\left(U^2\!\cdot\partial_{U^1}\right)^{s_2\!-\!3\!-\!\Delta}\,\hat{E}^2_{\{s_1\!+\!s_2\!-\!\Delta\!-\!3,\Delta\!+\!2\}}\,,
\ee
where \mt{-2\leq\Delta \leq s_2-3}\,.
\item \mt{n=s_1+s_2-\Delta-2} 
\be
\label{n2}
E^2=\left(U^2\!\cdot\partial_{U^1}\right)^{\Delta\!+\!1\!-\!s_1}\,\hat{E}^2_{\{\Delta\!+\!1,s_1\!+\!s_2\!-\!\Delta\!-\!2\}}\,,
\ee
where \mt{s_1-1\leq\Delta\leq s_1+s_2-2}\,.
\end{itemize}
Moreover, the gauge parameter \mt{E^1} is completely determined in terms of \mt{E^2} via eq.~\eqref{2famC}.\\
{\bf 2)} \mt{\Delta=s_1-2}\,. In this case, eqs.~(\ref{2famC}\,,\,\ref{2famC2}) imply 
\be
\label{s1-2}
E^1=\hat{E}^1_{\{s_1\!-\!1,s_2\}}\,,\qquad E^2=0\,.
\ee
Notice that, if \mt{s_1=s_2=s}\,, the YT associated to \mt{\hat{E}^1_{\{s\!-\!1,s\}}} is no longer admissible and therefore no gauge symmetry survives. 

All in all, the number of possible values of \mt{\Delta} allowing for gauge symmetries is \mt{2\,s_2\!+\!1}\,, for \mt{s_1\neq s_2} and \mt{2\,s}\,, for \mt{s_1=s_2=s}\,. For each of these values, only one irreducible gauge parameter, associated to one of the irreducible components of \mt{\Phi}\,, can be preserved.\footnote{In general, the number of irreducible gauge parameters associated to the \mt{s_2+1} (or \mt{s+1}) components of a reducible field with \mt{s_1>s_2} (or \mt{s_1=s_2=s}) is \mt{2\,s_2+1} (or \mt{2\,s})\,. The reason is that, except for the totally-symmetric component (and also for the rectangular one when \mt{s_1=s_2}) that possesses only one gauge parameter, all the others have two gauge parameters. However, only one of the two is associated to unitary AdS massless fields \cite{Metsaev:1995re}.} As a consequence, if one fixes the degree of homogeneity \mt{\Delta}\,, the corresponding reducible field will contain only one massless (unitary or non-unitary) irreducible component, while the remaining \mt{s_2} components will be massive or partially-massless. However, as already mentioned, due to the tangentiality constraint \eqref{t con E} these partially-massless fields do not have the corresponding gauge symmetries and we will not consider them in the following.  In order that the massless component be the one described by the YT \mt{\{s_1,s_2\}} \eqref{yng}, one can choose either \mt{\Delta=s_1-2} or \mt{\Delta=s_2-3}\,. In the former case, the corresponding gauge parameters are given by \eqref{s1-2}\,, while in the latter  case eqs.~(\ref{2famC}\,,\,\ref{n1}) lead to
\ba 
\Delta=s_2-3\,,\quad E^1&\!=\!&\tfrac{1}{s_2\!-\!s_1\!-\!1}\,U^2\!\cdot\partial_{U^1}\,\hat{E}^2_{\{s_1,s_2\!-\!1\}}\,,\nn
E^2&=&\hat{E}^2_{\{s_1,s_2\!-\!1\}}\,.
\ea
However, only the value \mt{\Delta=s_1-2} corresponds to unitary AdS massless fields \cite{Alkalaev:2009vm}. On the other hand, as explained right after eq.~\eqref{s1-2}\,, if \mt{s_1=s_2=s} this value of \mt{\Delta} preserves no gauge symmetry and the only possible choice is \mt{\Delta=s-3}\,.

\section{Two-familiy HS cubic interactions}\label{sec3}
In this section we describe how to translate the consistency condition for the TT parts of the interactions into a system of PDEs. For simplicity, we focus on couplings involving one mixed-symmetry gauge field (denoted by \mt{\Phi_1}) and two totally-symmetric gauge fields (denoted by \mt{\Phi_2} and \mt{\Phi_3}). Moreover, we also assume that \mt{\Phi_1} have \mt{s_1>s_2} and \mt{\Delta=s_1-2}\,, so that its irreducible component \mt{\{s_1,s_2\}} is a unitary massless field with gauge transformations given by \eqref{s1-2}
\be
\label{irrgt}
\delta^{\sst(0)}\,\Phi_1(X,U)=U^{{1}}\!\cdot\partial_X\,\hat{E}^1_{\{s_1\!-\!1,s_2\}}(X,U)\,.
\ee
The corresponding homogeneity conditions \eqref{h con} and \eqref{h con E}, written in operatorial form, read
\ba
\label{HTope}
&(X\cdot\partial_X-U^1\!\cdot\partial_{U^1}+2)\,\Phi_1(X,U)=0\,,\nn
&(X\cdot\partial_X-U^1\!\cdot\partial_{U^1})\,\hat{E}^1_{\{s_1\!-\!1,s_2\}}(X,U)=0\,.
\ea
Neglecting divergences, traces, on-shell vanishing terms and auxiliary fields, 
the most general expression for the TT parts of the cubic vertices reads \cite{Joung:2011ww,Joung:2012rv}
\ba
	\label{cubicact1}
	&&S^{\sst {(3)}}=
	\int d^{d+1}X\, \delta\Big(\sqrt{-X^2}-L\Big)\,
	C(Y,Z)\nn
	&&\times\,
	\Phi_1(X_1,U_1)\, \Phi_2(X_2,U_2)\, 
	\Phi_3(X_3,U_3)\, \Big|_{^{ X_i=X}_{ U_i=0}},
\ea
where \mt{C} is an arbitrary function of nine variables
\ba\label{Y and Z}
	&&\hspace{-1.1cm}Y^1_{1}=\partial_{U^1_{1}}\!\cdot\partial_{X_{2}}\,,\,\, Y^2_{1}=\partial_{U^2_{1}}\!\cdot\partial_{X_{2}}\,, \,\, Y_2=\partial_{U_{2}}\!\cdot\partial_{X_{3}}\,,\nn
	&&\hspace{-1.1cm} Y_3=\partial_{U_{3}}\!\cdot\partial_{X_{1}}\,, \,\, Z_1=\partial_{U_{2}}\!\!\cdot\partial_{U_{3}}\,, \,\, Z_2^{11}=\partial_{U_{3}^1}\!\!\cdot\partial_{U_{1}^1}\,,\nn
	&&\hspace{-1.1cm} Z_2^{12}=\partial_{U_{3}^1}\!\!\cdot\partial_{U_{1}^2}\,, \, Z_3^{11}=\partial_{U_{1}^1}\!\!\cdot\partial_{U_{2}^1}\,, \,Z_3^{21}=\partial_{U_{1}^2}\!\!\cdot\partial_{U_{2}^1}\,.
\ea
For instance, a vertex of the form
\be
\Phi_{\sst M_1M_2\,;\,N}\,
\partial^{\sst M_1}\,\partial^{\sst N}\,\Phi^{\sst P}\,\partial_{\sst P}\, \Phi^{\sst M_2}\,,
\ee
corresponds to the choice
\be
C=Y_1^1\,Y_1^2\,Y_2\,Z_2^{11}\,.
\ee
The insertion of the delta function in \eqref{cubicact1} is aimed at removing the ambiguities related to the divergent radial integral. As a consequence, total-derivative terms arising in the gauge variation no longer vanish, but give a contribution that can be cast in the form
\be
	\label{intdeltahat}
	\delta\Big(\sqrt{-X^2}-L\Big)\,\partial_{X^M}=
	-\,\delta\Big(\sqrt{-X^2}-L\Big)\,\tfrac{\hat\delta}{L}\,X_{\sst M}\,.
\ee
Here, the auxiliary variable \mt{\hat{\delta}} encodes the derivatives of the delta function according to the rule
\ba
	\delta^{\sst [n]}(R-L)&=&\left(\tfrac{1}{R}\,\tfrac{d}{dR}\right)^n\,\delta(R-L)\nn
	&\equiv&\delta(R-L)\,
	\left(-\tfrac{\hat\delta}L\right)^{n}\,.
\ea
In order to compensate these terms, the cubic vertices are to be amended by additional total-derivative contributions. These terms contain a lower number of derivatives
compared to the initial vertices and are weighted by proper powers of \mt{L^{-2}}, or equivalently, of \mt{\hat{\delta}/L}\,. This is the ambient-space counterpart of 
what happens in the intrinsic formulation:
the replacement of ordinary derivatives by covariant ones requires 
the inclusion of additional lower-(covariant)derivative vertices
in the Lagrangian.

Whenever a massless field takes part in the interactions,
the corresponding vertices must be compatible with
its gauge symmetries. Following the Noether procedure, gauge consistency can be studied order by order
in the number of fields, and
at the cubic level it translates into the condition
\be
\label{noether}
\delta^{\sst (0)}\,S^{\sst (3)}
	\approx 0\,,
	\ee
where \mt{\approx} means equivalence modulo the free field equations, divergences and traces. In our notation, the requirement \eqref{noether} is equivalent to imposing 
\ba
\label{gaugeconscond1}
\left[\,C\left(Y,Z\right)
\,,\,U^1_1\!\cdot\partial_{X_1}\,\right]\Big|_{{ U_1=0}}&\approx & f(Y,Z)\,U_1^1\!\cdot\partial_{U_1^2}\Big|_{U_1=0}\,,\nn
\left[\,C\left(Y,Z\right)
\,,\,U_2\!\cdot\partial_{X_2}\,\right]\Big|_{{ U_2=0}}&\approx & 0\,,\nn
\left[\,C\left(Y,Z\right)
\,,\,U_3\!\cdot\partial_{X_3}\,\right]\Big|_{{ U_3=0}}&\approx& 0\,,
\ea
where the right-hand side of the first equation appears due to the irreducibility of the gauge parameter. Using the Leibniz rule together with the HT conditions (\ref{t con}\,,\,\ref{t con E}\,,\,\ref{HTope}) and the identity \eqref{intdeltahat}, one can recast eqs.~\eqref{gaugeconscond1} into a system of three PDEs
\ba
\label{system}
&&\hspace{-1.2cm}{\bf1)}\,\Big[Y_3\partial_{Z_2^{11}}-Y_2\partial_{Z_3^{11}}+\tfrac{\hat{\delta}}{2\,L}\,\Big(Y_1^2\partial_{Y_1^2}-\,2\,Y_2\partial_{Y_2}\nn&&\hspace{-0.7cm}+2\,Y_3\partial_{Y_3}+Z_2^{12}\partial_{Z_2^{12}}-Z_3^{21}\partial_{Z_3^{21}}\Big)\partial_{Y_1^1}\nn
&&+\,\hat{\delta}\,Z_3^{21}\partial_{Z_3^{11}}\partial_{Y_1^2}\Big]\,C(Y,Z)\nn
&&\hspace{-0.7cm}=\,\Big(Y_1^2\partial_{Y_1^1}+Z_2^{12}\partial_{Z_2^{12}}+Z_3^{21}\partial_{Z_3^{21}}\Big)\,f(Y,Z)\,,\nn
&&\hspace{-1.2cm}{\bf2)}\,\Big[Y_1^1\partial_{Z_3^{11}}+Y_1^2\partial_{Z_3^{21}}-Y_3\partial_{Z_1}\nn
&&\hspace{-0.7cm}+\,\tfrac{\hat{\delta}}{2\,L}\,\Big(2\,Y_1^1\partial_{Y_1^1}-2\,Y_3\partial_{Y_3}+Y_1^2\partial_{Y_1^2}\nn
&&\hspace{-0.7cm}-\,Z_2^{12}\partial_{Z_2^{12}}-Z_3^{21}\partial_{Z_3^{21}}\Big)\partial_{Y_2}\Big]\,C(Y,Z)=0\,,\nn
&&\hspace{-1.2cm}{\bf3)} \,\Big[Y_2\partial_{Z_1}-Y_1^1\partial_{Z_2^{11}}-Y_1^2\partial_{Z_2^{12}}\nn
&&\hspace{-0.7cm}+\,\tfrac{\hat{\delta}}{2\,L}\,\Big(2\,Y_2\partial_{Y_2}-2\,Y_1^1\partial_{Y_1^1}-Y_1^2\partial_{Y_1^2}\nn
&&\hspace{-0.7cm}+\,Z_3^{21}\partial_{Z_3^{21}}+Z_2^{12}\partial_{Z_2^{12}}\Big)\partial_{Y_3}\Big]\,C(Y,Z)=0\,.
\ea
Finding the general solutions to such a system is a non-trivial task and we leave this issue for a forthcoming paper \cite{Joung:2012}. However, for given spins of the fields, all possible \mt{C} and \mt{f} can be written as polynomials in the variables \eqref{Y and Z} with arbitrary coefficients. In this way, eqs.~\eqref{system} reduce to a system of linear equations for the coefficients, which can be easily solved, for instance, with  Mathematica. Besides the interactions of the massless irreducible component \mt{\{s_1,s_2\}}\,, these solutions contain also the vertices associated to the remaining \mt{s_2} ones \mt{\{s_1\!+\!s_2\!-\!n,n\}}\,, \mt{n=0,\ldots,s_2\!-\!1}\,. 
In order to single the former types of interactions, one has to further project the solutions onto the corresponding YT \mt{\{s_1,s_2\}}\,.

\subsection*{Example: \mt{\{2,1\}\!-\!1\!-\!1}}

As an example, let us consider the interactions between the simplest hook gauge field and two spin-1 gauge fields, \emph{i.e.} the triple \mt{\{2,1\}\!-\!1\!-\!1}\,. In this case, the polynomial ansatz for \mt{C} involves nine arbitrary coefficients 
\ba
\label{Cansatz}
C&=&a_1\,(Y_1^1)^2\,Y_1^2\,Y_2\,Y_3+a_2\,Y_1^1\,Y_1^2\,Y_3\,Z_3^{11}\nn
&+&a_3\,Y_1^1\,Y_1^2\,Y_2\,Z_2^{11}+a_4\,(Y_1^1)^2\,Y_3\,Z_3^{21}\nn
&+&a_5\,(Y_1^1)^2\,Y_2\,Z_2^{12}+a_6\,(Y_1^1)^2\,Y_1^2\,Z_1\nn
&+&a_7\,Y_1^2\,Z_2^{11}\,Z_3^{11}+a_8\,Y_1^1\,Z_2^{11}\,Z_3^{21}\nn
&+&a_9\,Y_1^1\,Z_2^{12}\,Z_3^{11}\,,
\ea
associated to the nine possible Lorentz-invariant TT couplings one can start with. Similarly, one can expand \mt{f} as
\ba
\label{ansatz f}
f&=&b_1\,(Y_1^1)^2\,Y_2\,Y_3+b_2\,Y_1^1\,Y_2\,Z_2^{11}+b_3\,(Y_1^1)^2\,Z_1\nn
&+&b_4\,Y_1^1\,Y_3\,Z_3^{11}+b_5\,Z_2^{11}\,Z_3^{11}\,.  
\ea
Plugging the ansatze (\ref{Cansatz}\,,\,\ref{ansatz f}) into eqs.~\eqref{system} and solving the resulting linear system, one can express all the unknown coefficients \mt{a_i} and \mt{b_i} in terms of three independent ones \mt{\tilde{a}_1}\,, \mt{\tilde{a}_2} and \mt{\tilde{a}_3}\,. 
Plugging these solutions back into \eqref{Cansatz}, one ends up with \mt{C= \tilde{a}_1\,C_1 + \tilde{a}_2\, C_2 + \tilde{a}_3\, C_3}\,, where
\ba
\label{Csol}
C_1&=&(Y_1^1)^2\,Y_1^2\,Y_2\,Y_3+\tfrac{3}{2}\,\tfrac{\hat{\delta}}{L}\,(Y_1^1)^2\,Y_1^2\,Z_1\nn
C_2&=&(Y_1^1)^2\,Y_1^2\,Y_2\,Y_3-\tfrac{\hat{\delta}}{2\,L}\,\Big(2\,(Y_1^1)^2\,Y_3\,Z_3^{21}\nn
&+&3\,Y_1^1\,Y_1^2\,Y_2\,Z_2^{11}+Y_1^1\,Y_1^2\,Y_3\,Z_3^{11}\Big)\nn
&+&\tfrac{3}{4}\,\left(\tfrac{\hat{\delta}}{L}\right)^2\,\Big(Y_1^2\,Z_2^{11}\,Z_3^{11}+2\,Y_1^1\,Z_2^{11}\,Z_3^{21}\Big)\nn
C_3&=&(Y_1^1)^2\,Y_3\,Z_3^{21}-(Y_1^1)^2\,Y_2\,Z_2^{12}\nn
&+&Y_1^1\,Y_1^2\,\left(Y_2\,Z_2^{11}-Y_3\,Z_3^{11}\right)\nn
&+&\tfrac{3}{2}\,\tfrac{\hat{\delta}}{L}\,Y_1^1\,\Big(Z_2^{12}\,Z_3^{11}-Z_2^{11}\,Z_3^{21}\Big)\,.
\ea   
Finally, acting on the solutions \eqref{Csol} with the projector onto the hook YT \mt{\{2,1\}}
\be
Y_{\{2,1\}}=1-\tfrac{1}{3}\,U^2\!\cdot\partial_{U^1}U^1\!\cdot\partial_{U^2}\,,
\ee
one can select the \mt{3}-massless \mt{\{2,1\}\!-\!1\!-\!1} couplings. As a result, the coupling \mt{C_1} vanishes, while the remaining \mt{C_2} and \mt{C_3} both give 
 \ba
 \label{finresult}
C^{\sst \{2,1\}\!-\!1\!-\!1}&=&(Y_1^1)^2\,(Y_2\,Z_2^{12}-Y_3\,Z_3^{21})\nn
&+&Y_1^1\,Y_1^2\,(Y_3\,Z_3^{11}-Y_2\,Z_2^{11})\nn
&+&\tfrac{3}{2}\,\tfrac{\hat{\delta}}{L}\,Y_1^1\,(Z_2^{11}\,Z_3^{21}-Z_2^{12}\,Z_3^{11})\,.
\ea
Hence in the end one is left with only one consistent coupling involving a three-derivative part which is 
dressed by a one-derivative AdS tail.


\section*{Acknowledgement}
I am very grateful to
E. Joung and M. Taronna for collaborations on the topics presented in this note. I would like to thank also A. Sagnotti for reading the manuscript and for useful suggestions. Finally, I thank E.D. Skvortsov and Y.M. Zinoviev for discussions.  
The present research was supported in part by Scuola Normale Superiore, by INFN (I.S. TV12) and by the MIUR-PRIN contract 2009-KHZKRX.

\end{multicols}

\begin{thebibliography}{99}
\bibitem{Bekaert:2010hw} Bekaert X. and Boulanger N. and Sundell P. 2010  [arXiv:1007.0435] 
\bibitem{Sagnotti:2011qp} Sagnotti A. 2011  [arXiv:1112.4285] 
\bibitem{Campoleoni:2008jq} Campoleoni A. and Francia D. and Mourad J. and Sagnotti A. 2009 {\it Nucl.Phys.} {\bf B815} 289-367\,; 2010 {\it Nucl.Phys.} {\bf B828} 405-514                       
\bibitem{Vasiliev:1988sa} Vasiliev M.A. 1989 {\it Annals Phys.} {\bf 190} 59-106\,; 2003 {\it Phys.Lett.} {\bf B567} 139-151 
\bibitem{Zinoviev:2002ye} Zinoviev Y.M. 2002  [hep-th/0211233] 
\bibitem{deMedeiros:2003px} de Medeiros P. 2004 {\it Class.Quant.Grav.} {\bf 21} 2571-2593  
\bibitem{Burdik:2011cb} Burdik C. and Reshetnyak A. 2012 {\it J.Phys.Conf.Ser.} {\bf 343} 012102  
\bibitem{Alkalaev:2005kw} Alkalaev K.B. and Shaynkman O.V. and Vasiliev M.A. 2005 {\it JHEP} {\bf 0508} 069  
\bibitem{Zinoviev:2009vy} Zinoviev, Y.M. 2009 {\it Nucl.Phys.} {\bf B821} 21-47\,; 2010 {\it Nucl.Phys.} {\bf B826} 490-510  
\bibitem{Alkalaev:2010af} Alkalaev K.B. 2011 {\it JHEP} {\bf 1103} 031  
\bibitem{Skvortsov:2009zu} Skvortsov E.D. 2009 {\it J.Phys.} {\bf A42} 385401  
\bibitem{Alkalaev:2009vm} Alkalaev K.B. and Grigoriev M. 2010 {\it Nucl.Phys.} {\bf B835} 197-220\,; 2011 {\it Nucl.Phys.} {\bf B853} 663-687  
\bibitem{Grigoriev:2012xg} Grigoriev M. 2012  [arXiv:1204.1793] 
\bibitem{Campoleoni:2012th} Campoleoni A. and Francia D. 2012 [arXiv:1206.5877]  
\bibitem{Zinoviev:2010av} Zinoviev Y.M. 2011 {\it JHEP} {\bf 1103} 082 
\bibitem{Boulanger:2011qt} Boulanger N. and Skvortsov E.D. and Zinoviev Y.M. 2011 {\it J.Phys.} {\bf A44} 415403 
\bibitem{Zinoviev:2011fv} Zinoviev Y.M. 2012 {\it Class.Quant.Grav.} {\bf 29} 015013
\bibitem{Boulanger:2011se} Boulanger N. and Skvortsov E.D. 2011 {\it JHEP} {\bf 1109} 063 
\bibitem{Metsaev:1995re} Metsaev R.R. 1995 {\it Phys.Lett.} {\bf B354} B354\,; 1997 {\it Lect.Notes Phys.} {\bf 524} 331-340 
\bibitem{Brink:2000ag} Brink L. and Metsaev R.R. and Vasiliev M.A. 2000 {\it Nucl.Phys.} {\bf B586} 183-205
\bibitem{Boulanger:2008up} Boulanger N. and Iazeolla C. and Sundell P. 2009 {\it JHEP} {\bf 0907} 013\,; 2009 {\it JHEP} {\bf 0907} 014
\bibitem{Taronna:2011kt} Taronna M. 2012 {\it JHEP} {\bf 1204} 029
\bibitem{Joung:2011ww} Joung E. and Taronna M. 2011 {\it Nucl.Phys.} {\bf B861} 145-174
\bibitem{Joung:2012rv} Joung E. and Lopez L. and Taronna M. 2012 {\it JHEP} {\bf 1207} 041\,; 2012  [arXiv:1207.5520]
\bibitem{Fradkin:1986qy} Fradkin E.S. and Vasiliev M.A. 1987 {\it Nucl.Phys.} {\bf B291} 141\,; 1987 {\it Phys.Lett.} {\bf B189} 89-95
\bibitem{Boulanger:2008tg} Boulanger N. and Leclercq S. and Sundell P. 2008 {\it JHEP} {\bf 0808} 056
\bibitem{Fronsdal:1978vb} Fronsdal C. 1979 {\it Phys.Rev.} {\bf D20} 848-856
\bibitem{Biswas:2002nk} Biswas T. and Siegel W. 2002 {\it JHEP} {\bf 0207} 005
\bibitem{Deser:1983mm} Deser S. and Nepomechie R.I. 1984 {\it Annals Phys.} {\bf 154} 396\,; 1983 {\it Phys.Lett.} {\bf B132} 321
\bibitem{Deser:2001pe} Deser S. and Waldron A. 2001 {\it Phys.Rev.Lett.} {\bf 87} 031601\,; 2001 {\it Phys.Lett.} {\bf B508} 347-353\,; 2001 {\it Nucl.Phys.} {\bf B607} 577-604\,; 2001 {\it Phys.Lett.} {\bf B501} 134-139\,; 2001 {\it Phys.Lett.} {\bf B513} 137-141\,; 2003 {\it Nucl.Phys.} {\bf B662} 379-392 
\bibitem{Skvortsov:2006at} Skvortsov E.D. and Vasiliev M.A. 2006 {\it Nucl.Phys.} {\bf B756} 117-147
\bibitem{Francia:2008hd} Francia D. and Mourad J. and Sagnotti A. 2008 {\it Nucl.Phys.} {\bf B804} 383-420
\bibitem{Joung:2012} Joung E. and Lopez L. and Taronna M. to appear 




\end{thebibliography}
\end{document}